\documentclass[prl,twocolumn,amsmath,amssymb]{revtex4} 

\usepackage{graphicx}
\usepackage{subfigure} 
\usepackage{dcolumn}
\usepackage{array} 
\usepackage{bm}
\usepackage[hang]{footmisc}

\usepackage[usenames,dvipsnames]{color}

\begin{document} 
 
\title{\bf{Hyperferroelectrics: proper ferroelectrics with persistent polarization} \\[11pt] } 
\author{Kevin F. Garrity, Karin M. Rabe and David Vanderbilt} 
\affiliation{Department of Physics and Astronomy\\ 
Rutgers University, Piscataway, NJ 08854} 
 
\date{\today} 
\begin{abstract} 
All known proper ferroelectrics are unable to polarize normal to a
surface or interface if the resulting depolarization field is
unscreened, but there is no fundamental principle that enforces this behavior.
In this work, we introduce hyperferroelectrics, a new class of proper ferroelectrics which
polarize even when the depolarization field is unscreened, this condition being
equivalent to instability of a longitudinal optic mode in addition to the transverse-optic-mode 
instability characteristic of proper ferroelectrics.  We use
first principles calculations to show that several recently discovered
hexagonal ferroelectric semiconductors have this property, and we
examine its consequences both in the bulk and in a superlattice
geometry.
\end{abstract} 
 
 
\maketitle 
 
\marginparwidth 2.7in 
\marginparsep 0.5in 
\def\dvm#1{\marginpar{\small DV: #1}} 
\def\kfg#1{\marginpar{\small KFG: #1}} 
\def\kmr#1{\marginpar{\small KMR: #1}} 
\def\scr{\scriptsize} 

Ferroelectrics, which are materials with a non-zero spontaneous polarization that
can be switched by an external electric field, have been extensively
studied both experimentally and theoretically.
Much of the work on ferroelectrics has focused on proper
ferroelectrics, such as BaTiO$_3$.
These have a non-polar reference structure
that is related to the ferroelectric ground state by a
polar distortion that lowers the energy
in zero macroscopic electric field, 
corresponding to an unstable transverse optic (TO) mode.
However, a slab of a typical proper displacive ferroelectric with
insulating surfaces will not spontaneously polarize with polarization
normal to the surface, because at quadratic order in the polarization the energetic cost of the resulting
depolarization field is larger than the energy gain from freezing in
the distortion~\cite{loto}.  In order to polarize, the
depolarization field must be screened, as for example by a metallic
electrode placed on the surfaces of the
ferroelectric slab~\cite{ferrothinfilm}.

In contrast to proper ferroelectrics, improper ferroelectrics do not
have an unstable polar distortion in their high-symmetry structure.
Instead, these materials have one or more unstable non-polar
distortions. However, when these distortions assume non-zero values
they break inversion symmetry in the material, resulting in a non-zero
polarization~\cite{improper,improper_sto,improper_perov,benedek}.
Because the primary energy-lowering distortion in an improper
ferroelectric is non-polar,
the depolarization field is too weak to prevent
the instability.  Thus, a slab cut from
such a material can develop a non-zero polarization normal to
the surface~\cite{thinimproper}.

In this work, we demonstrate a new class of ``hyperferroelectrics.''
These are proper 
ferroelectrics in which
the polarization persists in the presence of
a depolarization field.
Using first-principles calculations, we identify hyperferroelectrics in the
recently discovered class of
hexagonal $ABC$ semiconducting ferroelectrics
~\cite{abcferro}.
Using first-principles-based modeling, we show that
hyperferroelectrics have an electric equation of state that is
qualitatively different from those of both proper and improper
ferroelectrics, resulting in persistent polarization regardless of
screening and unique dielectric behavior.
Finally, we discuss the potential applications of hyperferroelectrics,
whose ability to polarize in ultra-thin layers may allow
the creation of
highly tunable thin-film or superlattice structures displaying
ultra-fast switching behavior.

We perform first-principles density functional theory (DFT)
calculations~\cite{hk,ks} within the local-density approximation~\cite{lda1} 
using the Quantum Espresso code~\cite{QE}.  
We use
ultrasoft~\cite{ultrasoft} pseudopotentials from the GBRV
high-throughput pseudopotential set~\cite{GBRV,GBRVwebsite}.
Phonon frequencies, Born effective charges, and electronic dielectric constants are calculated using DFT perturbation
theory~\cite{dftpt1,dftpt2,dftpt3}, and polarization is calculated
using the Berry phase method~\cite{berry}.

We begin by reviewing the properties of normal proper ferroelectric
materials, which in their high-symmetry phase have at least one unstable TO mode, specifically,
a $\Gamma$ mode that is unstable under
zero macroscopic electric field ($\mathcal{E}\!=\!0$) boundary conditions.
The frequency of this mode can be obtained
from first-principles computation of the force-constant matrix with the usual periodic boundary conditions.
The longitudinal optic (LO) modes can then be obtained by adding to the
force-constant matrix a non-analytic long-range Coulomb term that
schematically takes the form $(Z^*)^2 / \epsilon^\infty$,
where $Z^*$ are the Born effective charges and
$\epsilon^\infty$ is the electronic contribution to the dielectric
constant, generating the well-known LO-TO splitting~\cite{ashcroft}.  For
normal proper ferroelectrics, this non-analytic term is sufficiently
large that all the LO polar modes are stable; in other words, the
depolarization field resulting from the long-range Coulomb interaction
will prevent the ferroelectric from polarizing under fixed $D\!=\!0$
boundary conditions.
For typical perovskite oxides, the
strength of the depolarization field must be weakened by at least 90\%
to allow for a non-zero polarization with $D\!=\!0$~\cite{loto}.

\begin{figure} 
\centering 
\includegraphics[width=3.5in]{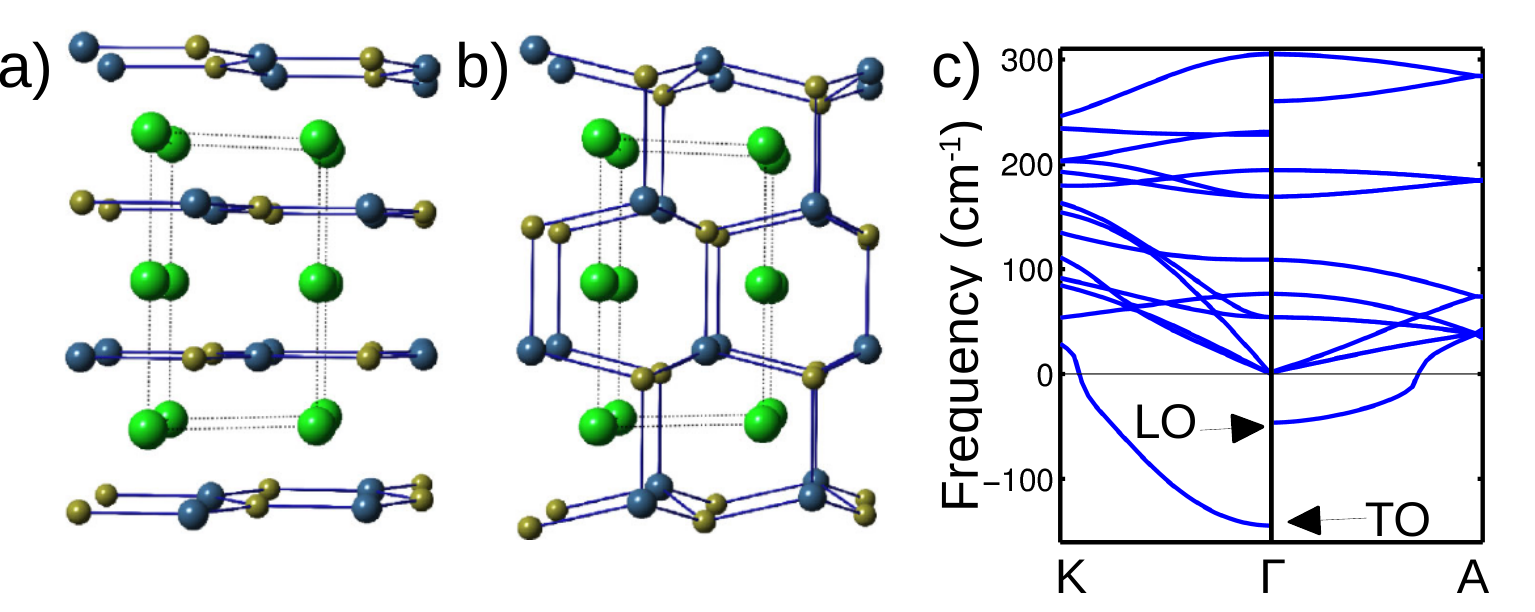} 
\caption{Structures of a) high-symmetry ($P6_3/mmc$) and b) polar
($P6_3/mc$) $ABC$ ferroelectrics.  The large green atom is the
`stuffing' atom.  c) Phonon spectrum of high-symmetry LiBeSb, from K
$(\pi/3a,\pi/3a,0)$ to $\Gamma (0,0,0)$ to A
$(0,0,\pi/2c)$ (imaginary frequencies are plotted as negative
numbers).}
\label{fig:phonon} 
\end{figure} 

While large-band-gap oxide ferroelectrics, which typically have large
$Z^*$'s and small $\epsilon^\infty$'s, have all LO modes stable, there is
no fundamental principle that enforces this stability.  
In fact, as we demonstrate in detail below, unstable LO modes can be found 
in semiconducting hexagonal $ABC$
ferroelectrics. 
The crystal structure is
shown in Fig.~\ref{fig:phonon}(a-b) (space group $P6_3mc$, LiGaGe structure type).
The high-symmetry phase of these materials consists of layers of two atoms in an
$sp^2$-bonded honeycomb lattice separated by layers of a third
`stuffing' atom, as shown in Fig.~\ref{fig:phonon}(a-b).  The polar
phase is reached by a single $\Gamma^{2-}$ phonon mode, which consists
primarily of a buckling in the honeycomb layers as the atoms move from
an $sp^2$ environment towards $sp^3$ bonding, resulting in polarization in the $z$ direction~\cite{abcferro}.  

In Table~\ref{tab:bulk}, we report the lowest TO and LO phonon
frequencies, dielectric constants, as well as band gaps, $\Delta
Z^*_{zz}\!=\!\sqrt{\sum_m (Z^*_{zz})^2_m / N}$, 
and polarizations for a variety of $ABC$ ferroelectrics; those with
imaginary LO frequencies are, by definition, hyperferroelectrics.  The
relatively small $\Delta Z^*_{zz}\!\approx\!3$ and large
$(\epsilon^\infty)_{zz}\!\approx\!10-20$ both contribute to the weak
depolarization fields in these materials (for reference, cubic
perovskites typically have $\Delta Z^*_{zz}\!\approx\!5$ and
$(\epsilon^\infty)_{zz}\!\approx\!  6$).  Both the small effective
charges and large dielectric constants of $ABC$ ferroelectrics are
consequences of the covalent bonding and resulting small band gaps of
these semiconductors.  In Fig.~\ref{fig:phonon}(c), we show the phonon
dispersion for the hyperferroelectric LiBeSb, which is a previously
synthesized material~\cite{libesb1,libesb2}.  We
see that the lowest frequency phonon mode for $q\!\rightarrow\!0$ is
unstable regardless of the direction from which $\Gamma$ is
approached.

\begin{table} 
\begin{center} 
\begin{ruledtabular} 
\begin{tabular}{lcccccccccc} 
$ABC$ & $\omega_{\rm TO}$ & $\omega_{\rm LO}$ & 
   $(\epsilon^\infty)_{zz}$& $\Delta Z^*_{zz}$ & Gap & $P^{(\mathcal{E}=0)}$ & $P^{(D=0)}$ \\
      & (cm$^{-1})$  & (cm$^{-1}$)  &       &                   & (eV) & (C/m$^2$) & (C/m$^2$) \\
\hline
LiZnP  & 134$i$ & \;49     &  13.3 & 3.0 & 1.27 &  0.80 & 0  \\
NaMgP  & 131$i$ & 150      &  10.6 & 2.9 & 0.89 &  0.52 & 0  \\
\hline			     		 						      								      
LiZnAs & 118$i$ &\; 68$i$  &  15.5 & 3.0 & 0.48 &  0.73 & 0.02 \\
LiBeSb & 144$i$ &\; 47$i$  &  19.9 & 2.9 & 0.93 &  0.59 & 0.02 \\
NaZnSb &\;42$i$ &\; 14$i$  &  10.2 & 2.0 & 0.69 &  0.51 & 0.01 \\
LiBeBi & 171$i$ &  132$i$  &  22.1 & 2.9 & 0.83 &  0.54 & 0.02 \\
\end{tabular} 
\end{ruledtabular} 
\caption{Properties of $ABC$ hexagonal ferroelectrics.  Compounds
are listed with the stuffing atom first.  First-principles results
for high-symmetry phase: $\omega_{\rm TO}$ and $\omega_{\rm LO}$
are frequencies of unstable polar modes approaching $\Gamma$
along $\hat{\bf q} = (100)$ and $(001)$ respectively; $L_c$
is defined in the text; $(\epsilon^\infty)_{zz}$ is the $zz$
electronic dielectric constant; $\Delta Z^*_{zz}$ is the RMS $zz$
Born effective charge; `Gap' is the band gap. $P^{(\mathcal{E}=0)}$ is
the first-principles polarization at $\mathcal{E}\!=\!0$.
$P^{(D=0)}$ is the polarization computed from the model of
Eqs.~(\ref{eqn:model}-\ref{eqn:model2}) at $D\!=\!0$.
}
\label{tab:bulk} 
\end{center} 
\end{table} 

In order to investigate the electric equation of state of hyperferroelectrics, 
we use a simple first-principles-based model.
We first define a dimensionless polar internal degree of freedom, $u$,
as the buckling of the honeycomb layer, which varies from zero in the
high symmetry structure to one in the polar structure at $\mathcal{E}\!=\!0$.  Then, we expand the free
energy up to second order in $\mathcal{E}$, and up to sixth order in
$u$, with the $\mathcal{E}\!=\!0$ polarization included up to first order in $u$,
\begin{eqnarray}\label{eqn:model}
F(u,\mathcal{E})  = -a u^2 + b u^4 + c u^6 - P_s u\,\mathcal{E}
- \frac{1}{2} \chi_e(u) \mathcal{E}^2,
\end{eqnarray}
where $F$ is the free energy, $\chi_e(u)\!=\!\epsilon^\infty(u) - 1$ is the
zero-field electronic susceptibility as a function of $u$, and $P_s$, $a$,
$b$, and $c$ are constants.
The polarization, $P$, is then
\begin{eqnarray}\label{eqn:model2}
P(u)  = -\frac{\partial F}{\partial \mathcal{E}} = P_s u  +
\chi_e(u) \mathcal{E},
\end{eqnarray}
which allows us to identify $P_s$ as the spontaneous polarization of
the ground-state structure at zero electric field ($u\!=\!1$,
$\mathcal{E}\!=\!0$), justifying the notation for this constant.  
We fit this model to our materials by running a
series of calculations with $\mathcal{E}\!=\!0$ and $u$ fixed between 0
and 1.1, allowing all of the other internal degrees of freedom as well
as the lattice vectors to relax.  In addition, we calculate $\epsilon^\infty(u)$
for each structure, which we fit to a cubic spline.
\begin{figure} 
\centering 
\includegraphics[width=3.5in]{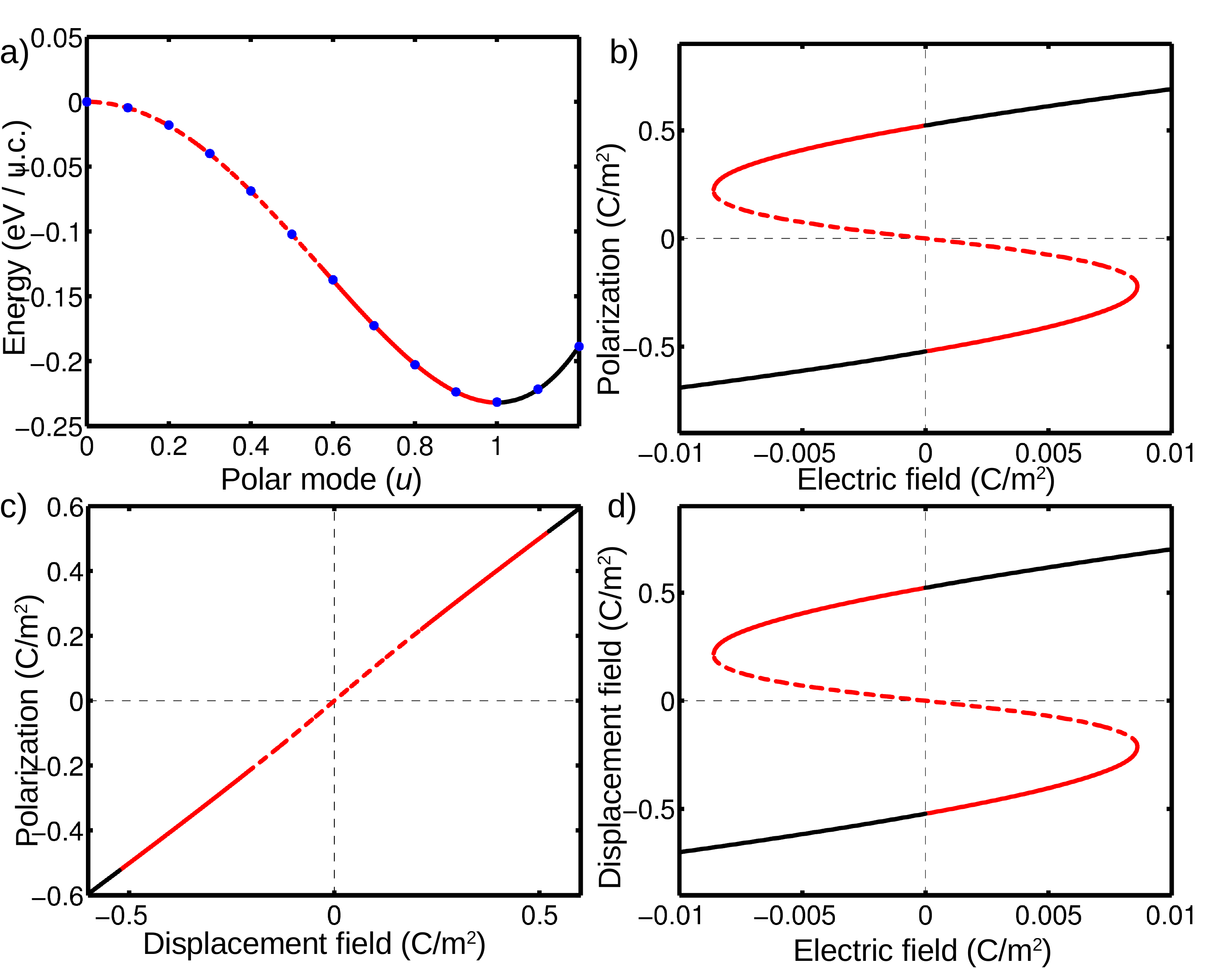} 
\caption{Computed energy landscape and electric equations of state for
normal ferroelectric NaMgP. a) Energy vs.\ polar mode
$u$. Dots are first principles; line is a fit to the model. b)
$P$ vs.\ $\epsilon_0 \mathcal{E}$.  c) $P$ vs.\ $D$. d) $D$
vs.\ $\epsilon_0 \mathcal{E}$.  Dashed red lines are locally unstable
at fixed $\mathcal{E}$; solid red lines are
locally stable; solid black lines are globally stable.}
\label{fig:namgp} 
\end{figure}

\begin{figure} 
\centering 
\includegraphics[width=3.5in]{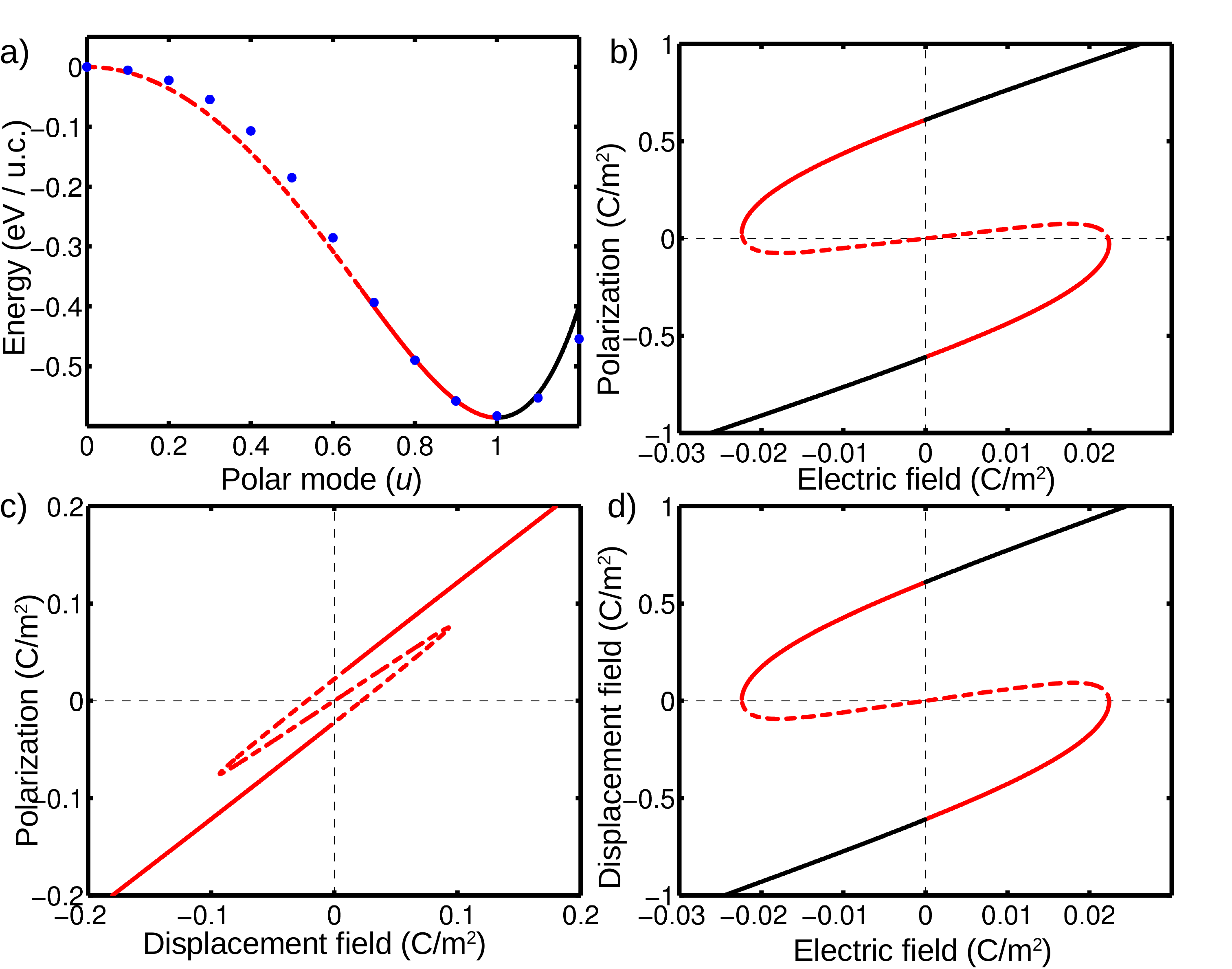} 
\caption{Computed energy landscape and electric equations of state for
hyperferroelectric LiBeSb.
Details as in Fig.~\ref{fig:namgp}.}
\label{fig:libesb} 
\end{figure} 

Using the model of Eqs.~(\ref{eqn:model}-\ref{eqn:model2}),
we can parametrically plot $\mathcal{E}(u)$, $P(u)$ and
$D(u)\!=\!\epsilon_0 \mathcal{E} + P$ versus each other, which we do
for both the normal ferroelectric NaMgP and the hyperferroelectric
LiBeSb in Figs.~\ref{fig:namgp}-\ref{fig:libesb}.  In both cases, we
indicate regions that are locally unstable, locally stable, and
globally stable under fixed-$\mathcal{E}$ boundary conditions.  In
locally unstable regions ($\partial P /\partial \mathcal{E}\!<\!0$),
the atomic degrees of freedom are at an unphysical maximum of the free
energy, rather than a minimum. 
In NaMgP, $P$ as a function of $\mathcal{E}$ is multi-valued at
$\mathcal{E}\!=\!0$, indicating that NaMgP is ferroelectric, with
spontaneous polarization as given in Table~\ref{tab:bulk}.  However,
$P$ vs. $D$ is single-valued, indicating that NaMgP will not polarize
under fixed $D\!=\!0$ boundary conditions, and thus is a normal proper
ferroelectric.  In contrast, for the hyperferroelectric LiBeSb, both
$P$ vs.\ $\mathcal{E}$ and $P$ vs.\ $D$ are multi-valued, so that the
material will spontaneously polarize under both fixed
$\mathcal{E}\!=\!0$ and fixed $D\!=\!0$ boundary conditions.  In
addition, the slope of $D$ vs.\ $\mathcal{E}$ indicates
$\epsilon^0\!=\!\partial D / \partial \mathcal{E}
|_{\mathcal{E}=0}\!>\!0$, despite the unstable polar mode.  As shown
in Table~\ref{tab:bulk} and Fig.~\ref{fig:libesb}(c), the $D\!=\!0$
polarization of hyperferroelectrics, $P^{(D=0)}$, which we compute
with the model of Eqs. \ref{eqn:model}-\ref{eqn:model2}, is small
compared to $P^{(\mathcal{E}=0)}$; however, the amplitude of the polar
mode remains surprisingly large.  The polar distortions of the
materials at $D\!=\!0$ are $25\!-\!75\%$ of their $\mathcal{E}\!=\!0$
values, but the resultant ionic polarization is largely canceled by the electronic
polarization $\chi_e(u) \mathcal{E}$ induced by the depolarization
field, resulting in a small net polarization.

\begin{figure} 
\centering 
\includegraphics[width=3.5in]{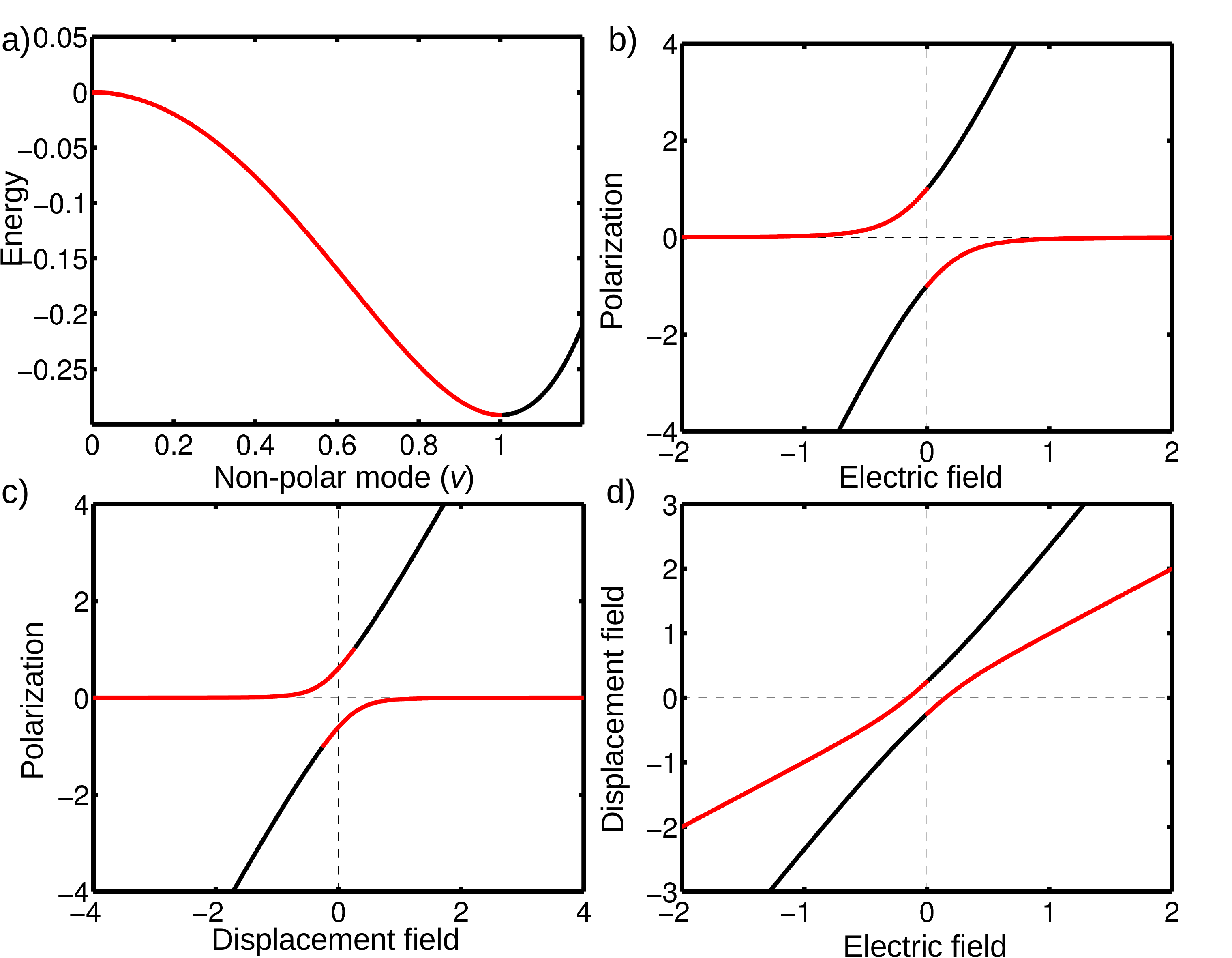} 
\caption{Energy landscape and electric equations of state for improper ferroelectric model of Eq.~(\ref{eqn:improper}).  a) Energy vs.\ non-polar mode
$v$. b) $P$ vs.\ $\mathcal{E}$. c) $P$
 vs.\ $D$. d) $D$ vs.\ $\mathcal{E}$.  All regions are locally stable
at
fixed $\mathcal{E}$; globally stable regions in black; other regions in red.}
\label{fig:improper} 
\end{figure} 

To emphasize the difference between hyperferroelectrics and improper
ferroelectrics, we briefly review a model of an improper
ferroelectric.  In the simplest improper ferroelectrics, the primary
order parameter, $v$, is non-polar, but it couples to a stable polar
mode $u$ with the form $u v^3$.  Then $u$, which appears only
to quadratic order, can be minimized over analytically, resulting in
an effective coupling between $v^3$ and $\mathcal{E}$:
\begin{eqnarray}\label{eqn:improper}
F(v,\mathcal{E})  = -a v^2 + b v^4  - c v^3\,\mathcal{E} - \frac{1}{2} \chi_e \mathcal{E}^2.
\end{eqnarray}
In Fig.~\ref{fig:improper}, we plot $P$ vs $D$ and $D$ vs
$\mathcal{E}$ for this model with typical parameters.  Similar to
hyperferroelectrics, improper ferroelectrics allow for a non-zero
polarization at $D\!=\!0$; however, the overall shape of the curves is
very different.  In particular, improper ferroelectrics lack a
structure with $D\!=\! P\!=\!0$.  This reflects the fact our model of
improper ferroelectrics always has a barrier to homogeneous switching
via external field ($\partial P /\partial\mathcal{E}\!>\!0$
everywhere), as that the effective coupling between the field and
third power of the non-polar distortion cannot overcome the primary
quadratic instability.  Clearly, then, the physical behavior of
improper ferroelectrics and hyperferroelectrics is qualitatively
different.

Returning to our main topic, we note that the unusual electric
properties of hyperferroelectrics mean that the ferroelectric phase
transition temperature, which decreases with decreasing screening,
will still be non-zero even under $D\!=\!0$ boundary conditions.  At
this temperature, $T_D$, the LO mode becomes unstable and the material
becomes a hyperferroelectric.  As a hyperferroelectric goes through
$T_D$ under $D\!=\!0$ boundary conditions, the inverse dielectric
constant will diverge, rather than the dielectric constant, which can
be understood by comparing the $D$ vs.\ $\mathcal{E}$ plots of normal
and hyperferroelectrics in Figs.~\ref{fig:namgp}(d) and
\ref{fig:libesb}(d), respectively.  In order to transition from the
normal to the hyperferroelectric state, the slope at the origin of
the $D$ vs.\ $\mathcal{E}$ curve, which is equal to the dielectric
constant, must pass through zero.

In order to demonstrate the consequences of the most notable quality
of hyperferroelectrics, their ability to polarize under fixed $D\!=\!0$
boundary conditions, we place our $ABC$ ferroelectrics in superlattice
configurations with thick slabs of non-polar $ABC$ materials.  We expect that normal ferroelectrics 
will not polarize in this geometry if there are no free charges, 
as a sufficiently thick non-polar layer will have $P\!=\!0$, which 
enforces $D\!=\!0$ boundary conditions on the ferroelectric, but 
hyperferroelectrics will still polarize under these conditions.

\begin{figure} 
\centering 
\includegraphics[width=3.5in]{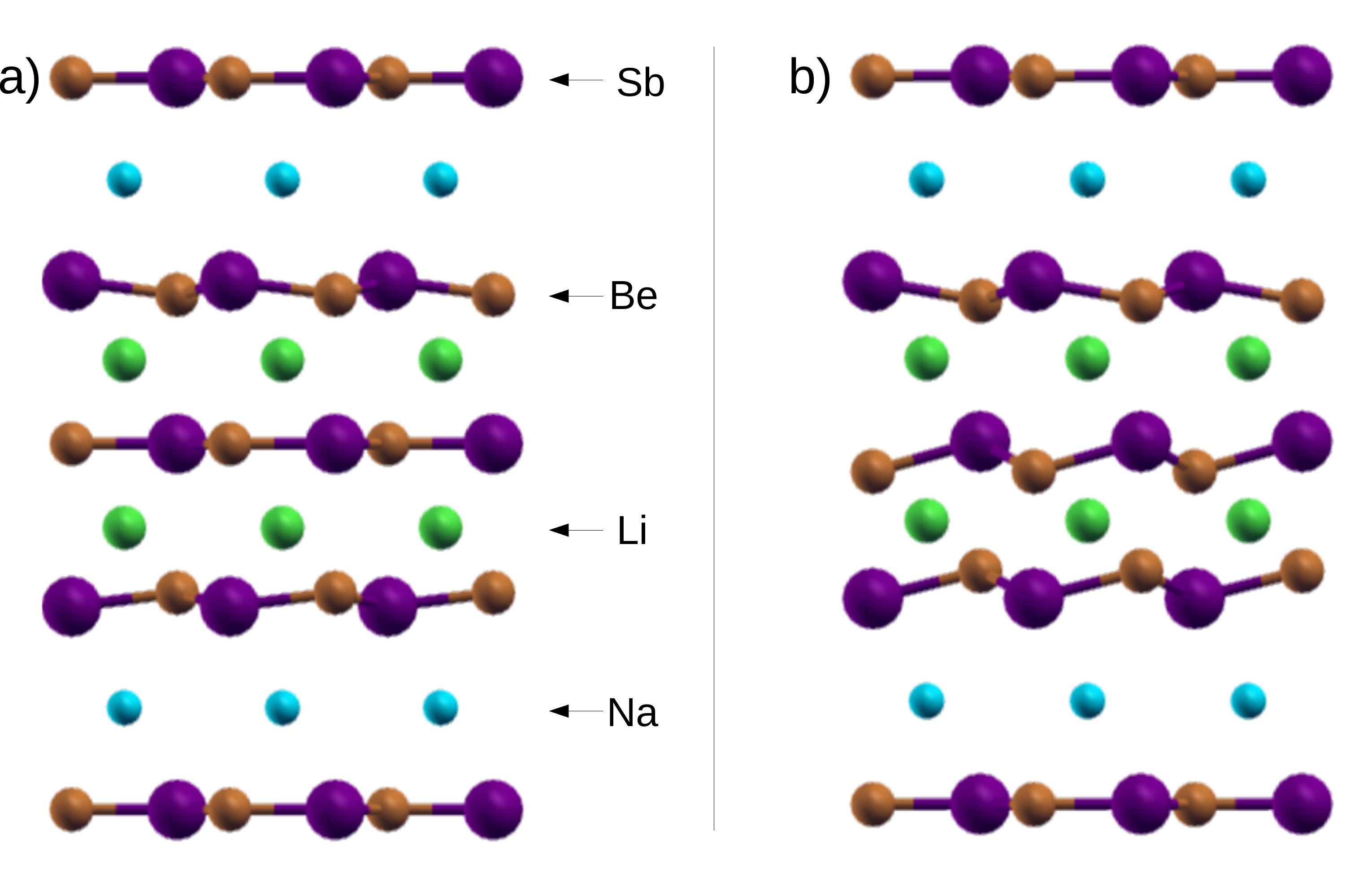} 
\caption{Interfacial region of a) non-polar and b) polar phases of 1/7 LiBeSb/NaBeSb superlattice.  The full supercell
has three additional unpolarized NaBeSb layers.}
\label{fig:superlat} 
\end{figure}

We consider superlattices consisting of ferroelectric $ABC$ materials
combined with non-polar hexagonal $ABC$ semiconductors, specifically normal ferroelectric NaMgP
with non-polar KZnSb and hyperferroelectric LiBeSb with non-polar NaBeSb, as shown in Fig.~\ref{fig:superlat}.  
We epitaxially
strain each superlattice to the in-plane lattice constant of the non-polar
material, allowing the $z$ lattice constant to relax.  

As shown in Table~\ref{tab:super}, the normal ferroelectric NaMgP has
essentially no polarization when in a superlattice with an insulating
material.  We attribute the tiny 10 meV energy lowering of the 1/7
NaMgP/KZnSb superlattice to interface effects, as the 
interfaces between NaMgP and KZnSb consist of single layers of NaZnSb, 
which as shown in Table~\ref{tab:bulk} is itself a hyperferroelectric.  
On the other hand, a
single polarized layer of the hyperferroelectric LiBeSb interfaced
with NaBeSb has a significantly lower energy and reduced band gap
relative to an unpolarized layer.  A second LiBeSb layer already
provides sufficient polarization to cause the system to become
metallic, due to dielectric breakdown, a field-induced overlap of conduction and valence bands leading
to charge transfer.

As already demonstrated, hyperferroelectrics can remain polarized down
to single atomic layers even when interfaced with normal insulators.
Such quasi-2d ferroelectric systems could have a variety of unusual
properties.  First, by adjusting the spacing of layers in a
superlattice, the polarization, well depth, band gap, and internal
electric field could all be tuned.  More speculatively, these
superlattice systems could display novel domain-wall motion or
super-fast switching behavior, as they consist of weakly-coupled
ferroelectric layers which may allow for easier domain nucleation, and
they support head-to-head and tail-to-tail domain walls.  Also, unlike
a normal ferroelectric, which requires asymmetric screening charges to
remain polarized, a hyperferroelectric can switch between two states
without the motion of screening charges between its surfaces or
interfaces, allowing hyperferroelectric slabs which are terminated by
vacuum or by non-polar insulators to be switched via an external
field.  In addition, in contrast to improper ferroelectrics, the
primary order parameter of hyperferroelectrics couples directly to an
applied electric field, which may allow for easier switching.
Finally, $ABC$ materials could be used to build an all-semiconducting ferroelectric
field effect transistor, side-stepping many of the materials
difficulties and interface effects that have hampered attempts to
interface ferroelectric oxides with
semiconductors~\cite{stoonsi,oxidesonsi,fpoxidesonsi}.

\begin{table} 
\begin{center} 
\begin{ruledtabular} 
\begin{tabular}{lcccccccc} 
Ferro. & Non-polar & Period & $\Delta E$  & Gap(HS) & Gap(FE) & $P^{(\mathcal{E}=0)}$\\
 & & & (eV) & (eV) & (eV) & C/m$^2$\\
\hline
NaMgP   &KZnSb &   1/7 &--0.01   & 0.32 & 0.35 & 0.007\\
NaMgP   &KZnSb &   2/6 &  0   & 0.69 & $-$ & 0\\
NaMgP   &KZnSb &   3/7 &  0   & 0.61 & $-$ & 0\\
NaMgP   &KZnSb &   4/6 &  0   & 0.68 & $-$ & 0\\
LiBeSb  &NaBeSb &  1/7 &--0.07   & 0.75 & 0.39 & 0.03\\
LiBeSb  &NaBeSb &  2/6 &--0.09   & 0.57 & $m$ & $m$\\
LiBeSb  &NaBeSb &  3/7 &--0.25   & 0.28 & $m$ & $m$\\
LiBeSb  &NaBeSb &  4/6 &--0.50   & 0.32 & $m$ & $m$\\
LiBeSb  &NaBeSb &  1/3 &--0.08   & 0.77 & 0.46 & 0.07\\
LiBeSb  &NaBeSb &  2/2 &--0.41   & 0.61 & 1.02 & 0.56\\
\end{tabular} 
\end{ruledtabular} 
\caption{Properties of superlattices. An $n/m$ ABC/A$'$B$'$C$'$ superlattice
consists of $n$ BC atomic layers separated
by A atomic layers, and $m$ B$'$C$'$ atomic
layers separated by A$'$ atomic layers, with A layers at both interfaces.  $\Delta E$ is the
energy gained by allowing a polar distortion.  `Gap(HS)' and `Gap(FE)' are
the band gaps for the non-polar and polar phases respectively; $m$
indicates a metal.  For insulators, $P^{(\mathcal{E}=0)}$ is the polarization 
for $\mathcal{E}\!=\!0$ boundary conditions.}
\label{tab:super} 
\end{center} 
\end{table} 

In conclusion, we have introduced a new class of ferroelectrics, hyperferroelectrics, 
and we have identified examples among the $ABC$ hexagonal semiconducting ferroelectric family.
These new ferroelectrics have a variety of interesting and potentially useful
properties, both in the bulk and as thin films.  Furthermore, this
work highlights the benefits of looking beyond well-studied materials systems
in the search for functional materials with novel properties.

\vspace{0.3cm} 
\noindent{\bf Acknowledgments} 
\vspace{0.3cm} 
 
This work was supported by ONR grants N0014-12-1-1035 and N0014-12-1-1040.
 

\end{document}